**Genetic diversity and fitness in small populations of partially asexual, self-incompatible plants**

Miguel Navascués[1,2], Solenn Stoeckel[3] and Stéphanie Mariette[4]

[1]Unidad de Genética, Centro de Investigación Forestal, INIA, Carretera de La Coruña km 7.5, 28040 Madrid, Spain

[2]Équipe Éco-évolution Mathématique, CNRS, UMR 7625 Écologie et Évolution, Université Pierre et Marie Curie & École Normale Supérieure, 46 rue d'Ulm 75230 Paris Cedex 05, France

[3]UMR 1099 BIO3P (Biology of Organisms and Populations applied to Plant Protection), INRA Agrocampus Rennes, F-35653 Le Rheu, France

[4]Unité de Recherche sur les Espèces Fruitières, INRA, Domaine de la Grande Ferrade, 71 avenue Edouard Bourlaux, BP 81, 33883 Villenave d'Ornon Cedex, France

Author for correspondence: Stéphanie Mariette, Tel: (33) 557 122 383, Fax: (33) 557 122 439, E-mail: smariett@bordeaux.inra.fr

**Keywords**

**Running title**: Asexuality and self-incompatibility in small populations

**Word count**: 5433

24    **Summary**

25    How self-incompatibility systems are maintained in plant populations is still a debated issue.

26    Theoretical models predict that self-incompatibility systems break down according to the

27    intensity of inbreeding depression and number of *S*-alleles. Other studies have explored the

28    role of asexual reproduction in the maintenance of self-incompatibility. However, the

29    population genetics of partially asexual, self-incompatible populations are poorly understood

30    and previous studies have failed to consider all possible effects of asexual reproduction or

31    could only speculate on those effects. In the present study, we investigated how partial

32    asexuality may affect genetic diversity at the *S*-locus and fitness in small self-incompatible

33    populations. A genetic model including an *S*-locus and a viability locus was developed to

34    perform forward simulations of the evolution of populations of various sizes. Drift combined

35    with partial asexuality produced a decrease in the number of alleles at the *S*-locus. In addition,

36    an excess of heterozygotes was present in the population, causing an increase in mutation

37    load. This heterozygote excess was enhanced by the self-incompatibility system in small

38    populations. In addition, in highly asexual populations, individuals produced asexually had

39    some fitness advantages over individuals produced sexually (due to the increased

40    heterozygosity, sex produces the homozygosis of deleterious alleles). Our results suggest that

41    future research on the role of asexuality for the maintenance of self-incompatibility will need

42    to (1) account for whole genome fitness (mutation load generated by asexuality, self-

43    incompatibility and drift) and (2) acknowledge that the maintenance of self-incompatibility

44    may not be independent of the maintenance of sex itself.

**Introduction**

Hermaphroditic plant species reproduce with variable rates of selfing, ranging from strict selfing to strict outcrossing (Barrett, 2002). Self-incompatibility (SI) is a reproductive system that prevents self-fertilization. In the case of heteromorphic self-incompatibility, distinct morphologies result in distinct compatibility groups, whereas in the case of homomorphic self-incompatibility, compatible individuals cannot be distinguished by their morphology (de Nettancourt, 1977). Most SI systems depend on physiological mechanisms that prevent pollen germination or pollen tube growth. In sporophytic self-incompatibility (SSI) systems, the compatibility of a pollen grain depends on the diploid genotype of the plant that produced it. In gametophytic self-incompatibility (GSI) systems, the compatibility of a pollen grain depends on its haploid genotype. GSI is more widespread than SSI (Glémin *et al.*, 2001).

Fisher (1941) showed that self-fertilization should have a selective advantage because a selfing genotype will transmit more copies of its genome than a non-selfing genotype (this has been termed the automatic advantage of selfing). However, numerous studies have shown that inbred offspring are less fit than outbred offspring. The relative decrease in the mean fitness of selfed versus outcrossed individuals (inbreeding depression) is generally recognized as the only main factor that counterbalances the selective advantage of selfing (Charlesworth and Charlesworth, 1987).

Consequently, the level of inbreeding depression in populations should play a determining role in the evolution of SI systems. Inbreeding depression decreases as population size becomes smaller due to reduced polymorphism for selection to act on (Bataillon and Kirkpatrick, 2000). Charlesworth and Charlesworth (1979) showed that the number of *S*-alleles also plays a role in maintaining SI, since a low number of alleles will limit the number of compatible crosses in the population. A decrease in population size can also cause a reduction in the number of *S*-alleles (Brennan *et al.*, 2003), and a self-compatible mutant can

70    be positively selected for (Reinartz and Les, 1994). Thus, small populations may be

71    particularly prone to the breakdown of SI due to weak inbreeding depression and low

72    numbers of *S*-alleles. However, small self-incompatible populations may maintain high levels

73    of inbreeding depression due to a sheltered load of deleterious alleles linked to the *S*-locus

74    (Glémin *et al*., 2001). The existence of a sheltered load has been demonstrated experimentally

75    in *Solanum carolinense* by Stone (2004) and Mena-Alí *et al*. (2009).

76    Overall, under a wide range of conditions, SI can evolve to self-compatibility. In effect, the

77    loss of SI systems is very frequent in plant evolution (Igic *et al*., 2008). However, the reasons

78    for which some species maintain a SI system while other species lose it are not completely

79    understood. It has been suggested that asexual reproduction, "when an individual produces

80    new individuals that are genetically identical to the ancestor at all loci in the genome, except

81    at those sites that have experienced somatic mutations" (de Meeûs *et al*., 2007), plays a role in

82    the maintenance or breakdown of SI. Two studies suggest that asexuality could relieve the

83    main selective pressures that favour the breakdown of SI. First, Chen *et al*. (1997) showed in

84    Australian Droseraceae that all self-incompatible taxa have effective forms of asexual

85    reproduction, whereas the obligatory sexual annual taxon, *Drosera glanduligera*, is self-

86    compatible. Their interpretation is that self-incompatible forms accumulate recessive lethal

87    polymorphisms, especially in association with biparental inbreeding generated by elevated

88    levels of asexual reproduction. The hypothetical high genetic load constitutes the selective

89    pressure that maintains the outbreeding mechanisms. Second, Vallejo-Marín and O'Brien

90    (2007) hypothesized that asexuality could provide reproductive assurance in cross-fertilizing

91    species subject to pollen limitation. They predicted that cross-fertilizing species subject to

92    pollen limitation would often have some means of asexual reprduction, and they found a

93    strong correlation between self-incompatibility and asexuality in *Solanum* (Solanaceae).

94    Conversely, Young *et al*. (2002) developed a contrasting hypothesis for the case of *Rutidosis*

*leiolepis* in which they found SSI and high rates of asexual reproduction. They suggested that increased asexual reproduction causes mate limitation by reducing genotype diversity at the *S*-locus, favouring a breakdown of self-incompatibility.

The net effect of asexual reproduction on the maintenance or breakdown of SI systems is therefore still unknown. Moreover, it is unclear to what extent the effects of asexual reproduction discussed by the various authors are actually present in partially asexual, self-incompatible populations. Theoretical studies have described the effects of partial asexuality on the inbreeding depression of self-compatible populations (Muirhead and Lande, 1997) and the number of *S*-alleles (Vallejo-Marín and Uyenoyama, 2008). However, no model has been developed so far to study the joint effect of asexuality, self-incompatibility and drift on the evolution of diversity and fitness parameters for a hermaphroditic species. Using individual-based simulations, we investigated the effect of (*i*) partial asexual reproduction and drift on diversity at the *S*-locus, (*ii*) combined asexuality, self-incompatibility and drift on the frequency of deleterious alleles. Thus, our main goal was to characterise the dynamics of two key factors (number of *S*-alleles and inbreeding depression) for the maintenance of SI in partially asexual, self-incompatible populations. It was not the aim of this work to estimate the probability of invasion of self-compatible alleles or other modifiers of reproduction.

**Materials and Methods**

*Genetic model*

The model considered in this study was based on the model developed by Glémin *et al.* (2001). This model consists of a population of $N$ (four population sizes evaluated: 25, 50, 100 and 1000) diploid hermaphroditic individuals with a GSI system. In addition, our model also included asexual reproduction at rate $c$ (probability that an individual is generated by asexual reproduction, seven values evaluated: 0, 0.5, 0.8, 0.9, 0.99, 0.999 and 1). We considered an asexual reproduction event as the production of a new independent individual that is an exact copy of the parental individual (or only different by somatic mutation) (e.g. de Meeûs *et al.*, 2007). As in Glémin *et al.* (2001), each individual genome possessed the *S*-locus, which regulated SI, and a viability locus, whose state determined individual fitness. Two neutral loci were also included for reference. The four loci were considered to be physically unlinked and were inherited independently through sexual reproduction.

Sexual reproduction events were controlled by the *S*-locus: crosses between individuals were only possible when the *S*-allele carried by the pollen grain was different from both *S*-alleles of the pistil (i.e. at least three different *S*-alleles were necessary in the population for sexual reproduction to occur). To focus on effects specifically attributable to asexuality on the maintenance and breakdown of SI, we assumed unlimited pollen availability and that sexual crosses were always fruitful. Thus, sexual reproduction was only limited when asexuality and genetic drift reduced the number of *S*-alleles to less than three within a population. In this case, independent of the original rate of asexual reproduction, all individuals reproduced asexually (*c*=1) until mutation introduced a third *S*-allele. Then, *c* was reset to its original value.

Individual fitness, *f*, was determined by the viability locus, which had two alleles *A* and *a*. Strength of the dominance of allele *A* over deleterious allele *a* was regulated by the dominance coefficient *h*, and strength of the selection against *a* was regulated by the selection coefficient *s*. Thus, relative fitness of genotypes were: $f_{AA}$=1, $f_{Aa}$=1-*hs* and $f_{aa}$=1-*s* (0<*s*<=1 and 0<*h*<= 1/2). Three selection regimes at the viability locus were considered: (*i*) a neutral case, to be able to study the effects of the interaction between SI and asexuality without the interference of selection, where parameters *s* and *h* were set to zero; (*ii*) a case of mildly deleterious and partially recessive mutations, for which values assigned to *s* and *h* were

144    respectively 0.1 and 0.2; and (*iii*) a case of highly recessive lethal mutations, values assigned

145    to $s$ and $h$ were respectively 1 and 0.02.

146    Mutations were allowed in all loci. Neutral loci mutated at rate $\mu_N=10^{-3}$, following a *k*-allele

147    model ($k=100$). Mutation at the *S*-locus followed an infinite-allele model (Kimura and Crow,

148    1964) with rate $\mu_S$; three values were considered for this rate ($\mu_S=10^{-3}$, $10^{-4}$ and $10^{-5}$ as in

149    Glémin *et al*., 2001). Mutations at the viability locus occurred at rate $\mu_1=10^{-3}$ from *A* to *a* and

150    at rate $\mu_2=10^{-4}$ from *a* to *A*.

151

152    *Individual-based simulations*

153    Forward-time simulations of a population of *N* diploid individuals with no overlapping

154    generations were performed. Simulations started with all individuals carrying two unique

155    alleles at the *S*-locus ($2N$ different *S*-alleles in the whole population), alleles *A* or *a* randomly

156    assigned (with equal probability, 0.5) to the viability locus and any of the *k* alleles ($k=100$)

157    randomly assigned (with equal probability, $1/k$) to the two neutral loci.

158    At each generation, the number of individuals generated by asexual reproduction, *x*, was

159    drawn from a Poisson distribution with mean $Nc$, and the remaining *N-x* individuals were

160    generated by sexual reproduction (parents contributing to these two types of offspring were

161    not determined in this step). Genotypes of the *N* individuals were generated one by one. If a

162    given individual originated by asexual reproduction, a genotype *i* was randomly drawn from

163    all individuals of the previous generation. That genotype was assigned to the individual with a

164    probability equal to the fitness of genotype *i*; this will be called the selection step. If the

165    genotype was not successfully assigned, the genotype draft and the selection step was

166    repeated until a genotype passed the selection step. If the individual originated by sexual

167    reproduction, the simulation followed a different procedure. First, an individual from previous

168    generation was randomly drawn to act as a mother and the ovule haploid genotype was

169 randomly generated. Then, a second individual was drawn to act as a father and the pollen

170 haploid genotype was randomly generated. If the *S*-allele at the pollen grain was not different

171 from both *S*-alleles of the mother (i.e. incompatible pollen), the pollen was discarded. New

172 fathers and pollen grain genotypes were generated until a compatible cross was obtained (i.e.

173 pollen availability was unlimited). Then, a diploid genotype *i* was generated from the haploid

174 genotypes of the ovule and the pollen. The selection step was carried out on genotype *i* as in

175 the case of asexual reproduction; if the genotype was not assigned the process was repeated

176 by drawing two new parents until a genotype was able to pass the selection step. Once all *N*

177 genotypes were selected, mutations were applied to each individual. The numbers of mutants

178 at the *S*-locus and at the two neutral loci were drawn from Poisson distributions with means

179 $2N\mu_S$ and $2N\mu_N$ respectively. At the viability locus, a mutant was formed with a probability $\mu_1$

180 if the allele was *A* and a probability $\mu_2$ if the allele was *a*. The rate of asexual reproduction

181 was a fixed parameter of our model and did not change with the selection process.

182

183 *Monitored genetic indices*

184 Simulations were run for 50 000 generations before beginning to monitor the genetic indices,

185 this allowed monitored indices to become stable and avoided the influence of initial

186 conditions in their values (Glémin *et al.*, 2001). From generation 50 000, indices were

187 calculated every 3 000 generations until 107 000 generations (i.e. 20 times). For the same

188 combination of parameters, 50 replicates of the simulations were performed providing 1 000

189 (indices calculated 20 times in each of the 50 replicates) evaluations of the monitored

190 variables for each set of parameter values. Indices were calculated from all genotypes of the

191 population (i.e. they were not estimates, but the true values of the population).

192 <u>Genetic diversity indices</u>: For each locus we calculated the number of alleles ($n_a$), allele

193 frequencies, effective number of alleles ($n_e = 1 \left/ \displaystyle\sum_{i=1}^{n_a} p_i^{\,2} \right.$, where $p_i$ was the frequency of allele),

194     expected heterozygosity ($H_e = 1 - 1/n_e$), observed heterozygosity ($H_o$, proportion of

195     heterozygous individuals) and the inbreeding coefficient ($F_{IS} = (H_e - H_o)/H_e$). Theoretical

196     predictions for some of these indices were obtained from the work of several authors for some

197     of the scenarios considered:

198     a) Vallejo-Marín and Uyenoyama (2008) studied the effects of partial asexuality in the

199     number of alleles of the $S$-locus using a diffusion approximation and found that the number of

200     common alleles ($n_c$, which is approximately equivalent to the effective number of $S$-alleles,

201     $n_e$, Takahata, 1990) is determined by:

202     (1)

$$1 = \theta \frac{2Mn_c}{2Mn_c - \theta(n_c-1)(n_c-2)} e^{\frac{2Mn_c}{(n_c-1)(n_c-2)} - \theta} \sqrt{\frac{2\pi(n_c-1)}{Mn_c + \theta(n_c-1)}} \left\{ \left[1 + \frac{\theta(n_c-1)}{Mn_c}\right] \frac{n_c-2}{n_c} \right\}^{\theta + Mn_c/(n_c-1)}$$

203     where $M = N(1-c)$ and $\theta = 2N\mu_S$. This equation was solved numerically to obtain the

204     expected effective number of $S$-alleles in the simulations without selection on the viability

205     locus.

206     b) Glémin $et\ al.$ (2001) obtained the expected $F_{IS}$ for a neutral locus linked to the $S$-locus as:

207     (2)

$$F_{IS} = -\frac{(1+2f_S)(n_c-1)}{1 + n_c^2 + 2f_S(n_c-1)(4Nr' + n_c-1)}$$

208     where $f_S$ is the scaling factor of the genealogy of the $S$-locus and $r'$ is the net recombination

209     rate between the two loci. For the case of partial asexuality, the number of common alleles

210     was given by Equation 1 and the scaling factor was $f_S = \dfrac{n_c^2(n_c-1)(n_c-2)}{4\theta[2Mn_c - \theta(n_c-1)(n_c-2)]}$

211     (Vallejo-Marín and Uyenoyama, 2008). To correct for the reduction of recombination due to

212     asexuality, we used $r' = r(1-c)$, where $r$ was the recombination rate during sexual

213     reproduction (i.e. $r$=0.5 for unlinked loci). This approach ignored mutation on the neutral

214     locus.

215   c) Finally, an expression of the expected $F_{IS}$ of a neutral locus in a self-compatible, partially

216   asexual population was derived following Balloux *et al.* (2003):

217   (3)     $F_{IS} = \dfrac{\gamma\left[(1-c)Ns-1\right]}{2N+\gamma\left[1+cN(s-2)-Ns\right]}$

218   where $\gamma = \left(1-\mu_N\right)^2$ and $s$ is the selfing rate. Selfing rate in Balloux *et al.* (2003) model is

219   defined as the rate of self-fertilization of individuals; thus a fertilization event between two

220   individuals from the same genet is considered an outcrossing event. Under random mating,

221   $s=1/N$ and for negligible mutation rates, $\gamma=1$.

222   <u>Linkage disequilibrium:</u> The linkage disequilibrium between pairs of loci ($S$-locus-viability

223   locus and the two neutral loci) was studied with the correlation coefficient between two loci,

224   $R_{GGD}$, developed by Garnier-Géré and Dillmann (1992) which is particularly appropriate for

225   partially asexual diploids (de Meeûs and Balloux, 2004).

226   <u>Fitness:</u> Mean fitness of the population ($\overline{W}$) was calculated from the genotype frequencies as

227   $\overline{W} = f(AA)\times 1 + f(Aa)\times(1-hs) + f(aa)\times(1-s)$, and the average fitness of individuals

228   potentially produced by selfing ($W_s$), outcrossing ($W_o$) and asexuality ($W_a$) were calculated as

229   $W_s = f(AA)\times 1 + f(Aa)\times\left[\frac{1}{4}\times 1 + \frac{1}{2}\times(1-hs) + \frac{1}{4}\times(1-s)\right] + f(aa)\times(1-s)$,

230   $W_o = 1\times p_A^2 + (1-hs)\times 2\times p_A\times p_a + (1-s)\times p_a^2$ (where $p_A$ and $p_a$ are the frequencies of alleles

231   $A$ and $a$) and $W_a = \overline{W}$. Inbreeding depression ($\delta$) and mutation load ($L$) were calculated as

232   $\delta = 1-(W_s/W_o)$ and $L = 1-\overline{W}$. In addition, we compared the fitness of individuals produced

233   by asexual reproduction ($W_a$) to the fitness of individuals produced sexually, using the ratio

234   $W_o/W_a$. $W_s$, $W_o$ and $W_a$ were calculated as the fitness of potential offspring if all individuals

235   were produced by selfing, by outcrossing and asexually, respectively. However, these modes

236   of reproduction might not be actually present in the different scenarios considered (e.g. in

237   self-incompatible populations, no selfing occurs, but potential $W_s$ and inbreeding depression

238   can be calculated).

239   <u>Fixation probabilities:</u> To interpret the results, we calculated the probabilities of four events:

240   fixation of alleles $A$ and $a$ in the viability locus, 'fixation' of viability locus heterozygote ($Aa$)

241   and fixation of a genotype in the $S$-locus (reduction of number of $S$-alleles to two, this was

242   only monitored in self-incompatible populations). These fixation probabilities were estimated

243   on 1 000 evaluations of the monitored variables as the proportion of times in which $p_a$=0,

244   $p_a$=1 and $H_o$=1 for the viability locus, or $n$=2 for the $S$-locus.

245

246   **Results**

247   *Effect of asexuality and drift on linkage disequilibrium between loci*

248   Probably the most obvious effect of partial asexual reproduction was the reduction of

249   recombination. Asexuality generated identical genotypes in offspring, so allele associations

250   were transmitted in the same way that they would be transmitted if they were physically

251   linked throughout the whole genome. In fully asexual populations, maximum linkage

252   disequilibrium was expected (equivalent to the linkage disequilibrium between fully linked

253   loci) and in partially asexual populations, recombination was reduced proportionally to the

254   rate of asexual reproduction. Figure 1 shows how non-random associations of alleles

255   increased with asexual reproduction rates. Linkage disequilibrium levels also depended on

256   population size because drift produced a departure from the expected values of frequencies of

257   allele combinations (Figure 1). Strong drift might have even caused some combinations of

258   alleles to be absent in the population.

259   In small populations, some generations with complete allelic association started to appear at

260   asexual reproduction rates of 0.8 and were predominant at rates of 0.99 and higher. In larger

261   populations, complete allelic association was only found in the fully asexual scenario. These associations of alleles were observed between loci subject to selection and between neutral loci, which showed that linkage disequilibrium was not a product of selective processes (Figure 1a and Figure 1b).

Physical linkage between the *S*-locus and loci with deleterious alleles can decrease the number of *S*-alleles (Uyenoyama, 2003). It can also dramatically increase inbreeding depression in small populations (Glémin *et al.*, 2001). Therefore, for the present model, it was important to discern whether the linkage disequilibrium generated by asexual reproduction is sufficient for the selective forces of the *S*-locus to interact with the viability locus. The results are presented below taking this perspective into account.

*Effect of asexuality and drift on diversity parameters: number of S-alleles and heterozygote excess*

The effective number of *S*-alleles decreased with asexuality (Figure 2) as shown analytically by Vallejo-Marín and Uyenoyama (2008). Balancing selection on the *S*-locus only occurred during sexual reproduction events. Since these events were scarce at high rates of asexual reproduction, the influence of balancing selection (which promotes high allelic diversity) on the population diminished. This can temporarily stop sexual reproduction by reducing the number of *S*-alleles to two (i.e. fixation of an *S*-locus genotype, Table 1).

No significant differences in the number of *S*-alleles were found among the three different selection regimes for the viability locus (frequency distributions of the number of *S*-alleles among different selection regimes were undistinguishable in a Kolmogorov-Smirnov test, *P*-value>0.98). It must be noted that small populations (with high linkage disequilibrium) were fixed for the wild type allele *A* most of the time (Table 2b) and this may explain the weak influence of deleterious alleles on the *S*-locus.

286    Another general effect of partial asexual reproduction in diploids was the reduction of allele

287    segregation. This can increase heterozygosity by the independent accumulation of mutations

288    on the two alleles of an asexual lineage (Pamilo, 1987; Birky, 1996). To quantify this effect,

289    we measured the inbreeding coefficient $F_{IS}$, which compares observed and expected

290    heterozygosities.

291    Theoretical equilibrium values for $F_{IS}$ at a neutral locus (unlinked to the $S$-locus) were very

292    similar in self-compatible and self-incompatible populations (Figure 3), and, for large

293    population sizes, the simulated populations followed the same pattern (Figure 3b). However,

294    for small population sizes, drift generated strong linkage disequilibrium between the $S$-locus

295    and other loci. Thus, balancing selection on the $S$-locus had a hitchhiking effect on other loci,

296    increasing their heterozygosity. This explained the contrasting $F_{IS}$ values on simulated self-

297    compatible and self-incompatible populations (Figure 3a). Analytical predictions do not

298    reflect this dramatic difference because they do not account for linkage disequilibrium. $F_{IS}$

299    decreased with asexuality and this decrease was enhanced by self-incompatibility in small

300    populations, particularly noticeable at asexual reproduction rates higher than 0.8. This

301    decrease in $F_{IS}$ observed in neutral loci also occurred at the viability locus (both for the mildly

302    deleterious and lethal recessive cases), which implied an increase in the frequency of the

303    deleterious allele (Figure 4).

304

305    *Effect of asexuality and drift on fitness*

306    For both self-compatible and self-incompatible populations, mutation load globally increased

307    with asexuality and population size, but a greater variance among generations and populations

308    was observed for very small populations (Figure 5). As mentioned above, the frequency of the

309    deleterious allele increased with asexuality in small self-incompatible populations, due to a

310    hitchhiking effect of the $S$-locus over the viability locus. In the self-incompatibility system,

when the population size was small enough, the mutation load increased with asexual reproduction rates; this load resulted from the increase of the deleterious allele frequency (Figure 5a and 5c). Under similar conditions in the self-compatible system, mutation loads were lower than the mutation load observed in the self-incompatible system (Figure 5b and 5d). In contrast, inbreeding depression had similar values in self-incompatible and self-compatible populations (Figure 6), showing some increases with asexual reproduction rate for lethal recessive mutations that was more apparent in large populations.

The average fitness was higher in sexual populations than in asexual populations due to Mendelian segregation (e.g. Figure 5; Chasnov, 2000; Kirkpatrick and Jenkins, 1989). The relative fitness between individuals produced by sexual and asexual reproduction was influenced very little by partial asexuality in a self-compatible population (Figure 7b and 7d). However, in small self-incompatible populations, the asexual reproduction rate increased the advantage of asexual compared to sexual reproduction (Figure 7a and 7c). These populations consisted mainly of heterozygous individuals (Figure 3). Therefore, asexual reproduction produced mainly *Aa* individuals, with only slightly lower fitness than the fittest *AA* haplotype. However, sexual reproduction produced a high proportion of *aa* individuals, reducing the average fitness of offspring, a reduction that was not compensated by the production of the fittest *AA* individuals. This could potentially lead to a positive feedback effect on the asexual reproduction rate, but this was not studied in our model where the asexual reproduction rate was a fixed parameter.

**Discussion**

The number of *S*-alleles and inbreeding depression are considered to be the main factors that influence the maintenance or breakdown of self-incompatibility systems (Charlesworth and Charlesworth, 1979). A decrease in the number of *S*-alleles is expected to favour the

336 breakdown of self-incompatibility since self-compatible mutants can invade the population,

337 whereas the maintenance of inbreeding depression within populations prevents the breakdown

338 of self-incompatibility. In the present study, we examined how drift and partial asexuality

339 may modify the dynamics of these two key parameters in a self-incompatible population.

340

341 *Drift, combined with partial asexuality, dramatically reduced the number of* S-*alleles*

342 As already predicted by Yokoyama and Hetherington (1982), we showed that the number of

343 *S*-alleles decreases with the effective size of the population. Thus, in small populations,

344 repeated bottlenecks or founder events may lead to the breakdown of the self-incompatibility

345 system. However, Karron (1987) found no differences in mating systems between rare and

346 widespread congeners across several families (but see also a study on some Brassicaceae

347 species, Kunin and Shmida, 1997). As for the impact of asexuality, our simulation study

348 confirmed a recent result demonstrated analytically by Vallejo-Marín and Uyenoyama (2008).

349 The number of *S*-alleles decreased with asexuality due to the weakened influence of balancing

350 selection in a partially asexual population (Figure 2). In contrast with other effects of

351 asexuality that were only observed for very high rates of asexuality, intermediate asexual

352 reproduction rates were sufficient to significantly reduce the number of *S*-alleles. High rates

353 of asexual reproduction combined with drift produced extremely low numbers of *S*-alleles.

354 This effect was strong enough to even reduce the number of *S*-alleles to two (i.e. fixation of

355 an *S*-locus genotype, Table 1), which stopped all possibility of sexual reproduction until a

356 new *S*-allele arose from mutation or migration. Therefore, in such extreme cases, this drastic

357 effect of asexuality on the number of *S*-alleles should favour the breakdown of self-

358 incompatibility. However, paradoxically, the number of *S*-alleles may not be very relevant in

359 a system where asexuality serves to provide reproductive assurance.

360

361     *Drift, combined with partial asexuality and self-incompatibility, led to an increase in*

362     *mutation load*

363     Drift influenced the level of mutation load. As population size decreased, the frequency of the

364     deleterious allele decreased and mutation load decreased, due to more effective purging. We

365     also observed a much greater variance in mutation load in very small populations (Table 2

366     and Figure 5, $N$=25). In this case, selection was overwhelmed by drift and this caused a higher

367     probability of fixation of the deleterious allele, and an increase in mutation load. Similar

368     results have been described by Bataillon and Kirkpatrick (2000), Glémin (2003) and Haag and

369     Roze (2007).

370     Drift is also a key component of mutation load in asexual populations (Haag and Roze, 2007).

371     Indeed, large asexual and sexual populations showed comparable frequencies of the

372     deleterious allele and comparable mutation loads (Figure 5; see also Haag and Roze, 2007).

373     However, mutation load was greater in small asexual populations than in small sexual ones.

374     This could be explained by the absence of segregation in asexual populations, in which

375     heterozygous individuals are present in high frequencies (Table 2 and Figure 5; see also Haag

376     and Roze, 2007).

377     In this study, we additionally showed that self-incompatibility played a role in increasing

378     mutation load in partially asexual, self-incompatible populations. Partial asexual reproduction

379     and small population sizes produced strong associations (i.e. linkage disequilibrium) between

380     deleterious alleles (on any locus of the genome) and *S*-alleles. Glémin *et al*. (2001) have

381     already shown that deleterious mutations on loci linked to the *S*-locus may be sheltered by

382     balancing selection acting on rare *S*-alleles. Asexuality and self-incompatibility both favoured

383     the increase of the heterozygote genotype *Aa* within the population, leading to an increase of

384     mutation load in small self-incompatible, partially asexual populations.

385

386    *Inbreeding depression remained low in small, partially asexual, self-incompatible*
387    *populations*

388    Inbreeding depression followed different patterns than the ones observed for mutation load.

389    For very small populations, inbreeding depression is expected to decrease (Bataillon and

390    Kirkpatrick, 2000), due to the absence of polymorphism at the viability locus. This was

391    observed in small self-compatible and self-incompatible populations (Glémin *et al.*, 2001, and

392    our own simulations, Figure 6). As for the impact of asexuality, our results showed no large

393    differences between the self-compatible and the self-incompatible cases for small populations.

394    The effect of strong drift, in the presence of strong asexuality, was to reduce genotype

395    diversity. In most cases, populations were fixed at one allele, but sometimes the population

396    was fixed at the heterozygote genotype *Aa*. This mainly occurred in self-incompatible

397    populations (see Table 2). With no or very low genotype diversity, inbreeding depression had

398    very low values, even if deleterious alleles were present at high frequencies in the population.

399    Chen *et al.* (1997) proposed that asexuality would favour the maintenance of SI. Their

400    reasoning was that asexual reproduction would increase a population's genetic load due to

401    lethal recessive mutations, which was confirmed by our simulation results. However,

402    mutation load and inbreeding depression did not show parallel trends, and the high level of

403    mutation load due to the combination of drift, asexuality and SI was accompanied by low

404    levels of inbreeding depression. Nevertheless, inbreeding depression generally tended to

405    increase with asexuality due to the accumulation of deleterious alleles and this should favour

406    the maintenance of SI.

407

408    *Evolution of partially asexual, self-incompatible populations: a third possible outcome*

409    Until now, the evolutionary interest of partially asexual, self-incompatible populations has

410    been focused on two possible outcomes: (1) the system is stable and SI is maintained or (2)

the system is unstable and SI disappears (Vallejo-Marín, 2007; Vallejo-Marín and O'Brien, 2007; Chen *et al*., 1997; Young *et al*., 2002). However, a third possible outcome should be considered: being the unstable system, sexual reproduction completely disappears. This type of event was observed in our simulations due to a reduction in the number of *S*-alleles to two (see above). However, we also studied how selection on the viability locus could contribute to the potential loss of sexuality. Our results showed that the fitness of individuals produced by asexual reproduction ($W_a$) was greater than that of individuals produced by random sexual mating ($W_o$) in small populations (Figure 7). Fitness of individuals (potentially) produced by selfing ($W_s$) was also lower than $W_a$ in small self-incompatible populations, suggesting that asexuality may have an immediate advantage over selfing under such circumstances.

Therefore, we speculate that asexual reproduction could take over sexual reproduction more easily than self-compatibility over self-incompatibility when low numbers of *S*-alleles reduce mate availability. Despite the effect of asexuality on inbreeding depression and *S*-locus diversity, asexual reproduction may offer an alternative means of reproduction under these adverse conditions and self-incompatibility may be temporarily dormant instead of breaking down. Asexual populations could subsequently recover a sexual self-incompatible reproductive system when new *S*-alleles are introduced in the population (for instance, by migration). However, prolonged periods of asexual reproduction may facilitate the accumulation of mutations affecting sexual traits, including self-incompatibility, since selective pressures are ineffective for these traits (Eckert, 2001). Thus, alternative periods of sexual and asexual reproduction may be necessary for such dynamics, which is probably the case for plants combining asexual and sexual reproduction (Bengtsson and Ceplitis, 2000). Nevertheless, the scenarios described here need to be confirmed by further and more complex simulations and other factors may favour self-compatible sexual reproduction, such as the more effective purging of deleterious mutations.

436

In this study, we used a model with physically unlinked loci and a single viability locus. However, because the linkage disequilibrium produced by asexuality affected the whole genome, population genomic models may be necessary to fully study the effect of deleterious alleles on the *S*-locus for small, partially asexual populations. In addition, physically linked loci with deleterious alleles (which were not studied here) may reduce the number of *S*-alleles (Uyenoyama, 2003) and the strength of that effect may be influenced by the rate of asexuality as discussed by Vallejo-Marín and Uyenoyama (2008). Moreover, in the present study, it was not possible to investigate the evolutionary dynamics of mating system and asexuality modifiers since these were assumed to be fixed parameters. It would be then interesting to develop a model to test the invasion propensity of self-compatible or asexual mutants. Finally, models incorporating migration or population structure are necessary to study more realistic dynamics of partially asexual, self-incompatible populations. Population structure is expected to affect the genetic diversity of viability and the *S*-locus, for instance by introducing a third allele via migration within subpopulations fixed at two *S*-alleles. These models should also take into account the specificities of the different types of asexual reproduction where, for instance, the asexual propagules may have a lower dispersal capacity (e.g. vegetative reproduction) and may create spatial clusters of individuals with the same genotype.

Recent theoretical work (Vallejo-Marín and Uyenoyama, 2008) and our own model lead to several predictions about *S*-locus diversity, mutation load and inbreeding depression within partially asexual self-incompatible populations. However, we need more experimental observations to test and validate these theoretical advances. First, as outlined by Vallejo-Marín and O'Brien (2007), if asexuality relieves selective pressures favouring the breakdown of SI, the co-occurrence of asexuality and self-incompatibility should be frequent. Association

461    of self-incompatibility and asexuality has been studied in only a few monospecific studies

462    (*Rutidosis leiolepis*, Young *et al*., 2002; *Eucalyptus morrisbyi*, Jones *et al*., 2005; *Prunus*

463    *avium*, Stoeckel *et al*., 2006). The association was found at a multispecific level by Chen *et*

464    *al*. (1997) in Australian Droseraceae and by Vallejo-Marín and O'Brien (2007) in Solanaceae.

465    More reports on species combining self-incompatibility and asexuality would enhance our

466    understanding of pressures explaining their maintenance. Second, experimental measurements

467    of asexual reproduction rates and number of *S*-alleles in species showing various levels of

468    asexual reproduction would be useful to test predictions from our simulations. Although

469    experimental measurements of inbreeding depression in self-incompatible species are

470    available (Mena-Alí *et al*., 2008), studies comparing inbreeding depression from different

471    populations with contrasting sizes and rates of asexual reproduction are lacking.

472

473    *Conclusions*

474    This study investigated for the first time the effect of partial asexual reproduction on the

475    fitness of self-incompatible populations. For loci unlinked to the S-locus and large

476    populations, fitness values were similar to those of a self-compatible population. However, in

477    small populations, the combination of drift, asexuality and SI increased mutation load due to

478    the accumulation of deleterious alleles. This increase in the mutation load was accompanied

479    by a selective advantage of asexually produced offspring compared to sexually produced

480    offspring. Therefore, future studies addressing the maintenance of SI in partially asexual

481    populations (by studying the invasibility of self-compatible genotypes) will need to (1)

482    account for whole genome fitness and (2) acknowledge that the process may not be

483    independent of the maintenance of sex itself (i.e. will need to establish whether modifiers of

484    asexual reproduction rates invade more easily than self-compatible alleles).

485

**Acknowledgements**

Collaboration between MN and SM was promoted through the implementation of the REPROFOR project, financed by the Spanish Ministry for Education and Science. We would like to thank Santiago C. González-Martínez and Ricardo Alía for helping to develop this collaboration. E. Porcher is acknowledged for helpful discussions on self-incompatibility evolution. We are also very thankful to P. Garnier-Géré for a detailed discussion on linkage disequilibrium estimators.

493 **References**

494 Balloux F, Lehmann L, de Meeûs T (2003). The population genetics of clonal and partially

495       clonal diploids. *Genetics* **164**: 1635-1644.

496 Barrett SCH (2002). The evolution of plant sexual diversity. *Nat Rev Genet* **3**: 274-284.

497 Bataillon T, Kirkpatrick M (2000). Inbreeding depression due to mildly deleterious mutations

498       in finite populations: size does matter. *Genet Res* **75**: 75-81.

499 Bengtsson BO, Ceplitis A (2000). The balance between sexual and asexual reproduction in

500       plants living in variable environments. *J Evolution Biol* **13**: 415-422.

501 Birky CW (1996). Heterozygosity, Heteromorphy, and Phylogenetic Trees in Asexual

502       Eukaryotes. *Genetics* **144**, 427-437.

503 Brennan AC, Harris SA, Hiscock SJ (2003). The population genetics of sporophytic self-

504       incompatibility in Senecio squalidus L. (Asteraceae): avoidance of mating constraints

505       imposed by low S-allele number. *Philos T R Soc B* **358**: 1047-1050.

506 Charlesworth D, Charlesworth B (1987). Inbreeding depression and its evolutionary

507       consequences. *Annu Rev EcolEvol S* **18**: 237-268.

508 Charlesworth D, Charlesworth B (1979). The evolution and breakdown of S-allele systems.

509       *Heredity* **43**: 41-55.

510 Chasnov JR (2000). Mutation-selection balance, dominance and the maintenance of sex.

511       *Genetics* **156**: 1419-1425.

512 Chen L, Stace HM, James SH (1997). Self-incompatibility, seed abortion and clonality in the

513       breeding systems of several Western Australian Drosera species (Droseraceae) . *Aust J*

514       *Bot* **45**: 191-201.

515 Eckert CG (2001). The loss of sex in clonal plants. *Evol Ecol* **15**: 501-520.

516 Fisher RA (1941). Average excess and average effect of a gene substitution. *Ann Eugen* **11**:

517       53-63.

518 Garnier-Géré P, Dillmann C (1992). A computer program for testing pairwise linkage
519          disequilibria in subdivided populations. *J Hered* **83**: 239.

520 Glémin S (2003). How are deleterious mutations purged? Drift versus nonrandom mating.
521          *Evolution* **57**: 2678—2687.

522 Glémin S, Bataillon T, Ronfort J, Mignot A, Olivieri I (2001). Inbreeding depression in small
523          populations of self-incompatible plants. *Genetics* **159**: 1217-1229.

524 Haag CR, Roze D (2007). Genetic load in sexual and asexual diploids: segregation,
525          dominance and genetic drift. *Genetics* **176**: 1663-1678.

526 Igic B, Lande R, Kohn JR (2008). Loss of self-incompatibility and its evolutionary
527          consequences. *Int J Plant Sci* **169**: 93-104.

528 Jones RC, McKinnon GE, Potts BM, Vaillancourt RE (2005). Genetic diversity and mating
529          system of an endangered tree Eucalyptus morrisbyi. *Aust J Bot* **53**: 367-377.

530 Karron JD (1987). A comparison of levels of genetic polymorphism and self-compatibility in
531          geographically restricted and widespread plant congeners. *Evol Ecol* **1**: 47-58.

532 Kimura M, Crow JF (1964). The number of alleles that can be maintained in a finite
533          population. *Genetics* **49**: 725-738.

534 Kirkpatrick M, Jenkins CD (1989). Genetic segregation and the maintenance of sexual
535          reproduction. *Nature*, **339**: 300-301.

536 Kunin WE, Shmida A (1997). Plant reproductive traits as a function of local, regional, and
537          global abundance. *Conserv Biol* **11**: 183-192.

538 de Meeûs T, Balloux F (2004). Clonal reproduction and linkage disequilibrium in diploids: a
539          simulation study. *Infect Genet Evol* **4**: 345—351.

540 de Meeûs T, Prugnolle F, Agnew P (2007). Asexual reproduction: genetics and evolutionary
541          aspects. *Cell Mol Life Sci* **64**: 1355-1372.

542    Mena-Alí J, Keser L, Stephenson A (2008). Inbreeding depression in *Solanum carolinense*
543         (Solanaceae), a species with a plastic self-incompatibility response. *BMC Evol Biol* **8**:
544         10.

545    Mena-Alí J, Keser L, Stephenson A (2009). The effect of sheltered load on reproduction in
546         *Solanum carolinense*, a species with variable self-incompatibility. *Sex Plant Reprod*
547         **22**: 63-71.

548    Muirhead CA, Lande R (1997). Inbreeding depression under joint selfing, outcrossing, and
549         asexuality. *Evolution* **51**: 1409-1415.

550    de Nettancourt D (1977). *Incompatibility in Angiosperms*, New York: Springer, Berlin
551         Heidelberg .

552    Pamilo P (1987). Heterozygosity in apomictic organisms. *Hereditas* **107**: 95-101.

553    Reinartz JA, Les DH (1994). Bottleneck-induced dissolution of self-incompatibility and
554         breeding system consequences in Aster furcatus (Asteraceae). *Am J Bot* **81**: 446-455.

555    Stoeckel S, Grange J, Fernandez-Manjarres JF, Bilger I, Frascaria-Lacoste N, Mariette S
556         (2006). Heterozygote excess in a self-incompatible and partially clonal forest tree
557         species - Prunus avium L. *Mol Ecol* **15**: 2109-2118.

558    Stone JL (2004). Sheltered load associated with S-alleles in *Solanum carolinense*. *Heredity*
559         **92**: 335-342.

560    Takahata N (1990). A simple genealogical structure of strongly balanced allelic lines and
561         trans-species evolution of polymorphism. *PNAS* **87**: 2419-2423.

562    Uyenoyama MK (2003). Genealogy-dependent variation in viability among self-
563         incompatibility genotypes. *Theor Popul Biol* **63**: 281-293.

564    Vallejo-Marín M (2007). The paradox of clonality and the evolution of self-incompatibility.
565         *Plant Signal Behav* **2**: 265-266.

566  Vallejo-Marín M, O'Brien HE (2007). Correlated evolution of self-incompatibility and clonal
567       reproduction in Solanum (Solanaceae). *New Phytol* **173**: 415-421.

568  Vallejo-Marín M, Uyenoyama MK (2008). On the evolutionary modification of self-
569       incompatibility: implications of partial clonality for allelic diversity and genealogical
570       structure. In *Self-Incompatibility in Flowering Plants. Evolution, Diversity, and*
571       *Mechanisms* . pp. 53-71.

572  Yokoyama S, Hetherington LE (1982). The expected number of self-incompatibility alleles in
573       finite plant populations. *Heredity* **48**: 299-303.

574  Young AG, Hill JH, Murray BG, Peakall R (2002). Breeding system, genetic diversity and
575       clonal structure in the sub-alpine forb *Rutidosis leiolepis* F. Muell. (Asteraceae). *Biol*
576       *Conserv* **106**: 71-78.

 **Figure legends**

 **Figure 1** Linkage disequilibrium ($R_{GDD}$ index, Garnier-Géré & Dillmann 1992) at increasing

579 rates of asexual reproduction. Box-plots represent median (black line), first and third quartiles

580 (box) and 5% and 95% percentiles (whiskers) of 1 000 observations from simulations

581 performed at seven rates of asexual reproduction ($c$=0, 0.5, 0.8, 0.9, 0.99, 0.999 and 1) and

582 four population sizes ($N$=25, 50, 100 and 1000). (a) $R_{GDD}$ index between $S$-locus and viability

583 locus ($s$=0.1, $h$=0.2); (b) $R_{GDD}$ index between two neutral loci.

584 **Figure 2** Effective number of $S$-alleles at increasing rates of asexual reproduction. Box-plots

585 represent median (black line), first and third quartiles (grey box) and 5% and 95% percentiles

586 (whiskers) of 1 000 observations from the simulations performed at six rates of asexuality

587 ($c$=0, 0.5, 0.8, 0.9, 0.99 and 1), population size $N$=25 (a) and $N$=100 (b), mutation rate $\mu_S$=10$^{-5}$,

588 $s$=0 and $h$=0 for the viability locus. Grey line represents the theoretical expected number of

589 common $S$-alleles obtained numerically from Equation 1 (Equation 3.7 in Vallejo-Marín &

590 Uyenoyama, 2008).

591 **Figure 3** Inbreeding coefficient $F_{IS}$ at a neutral locus at increasing rates of asexual

592 reproduction. Box-plots represent median (black line), first and third quartiles (white and grey

593 boxes, for self-compatible and self-incompatible populations respectively) and 5% and 95%

594 percentiles (whiskers) of 1 000 observations from the simulations performed at six rates of

595 asexuality ($c$=0, 0.5, 0.8, 0.9, 0.99 and 1), mutation rates $\mu_S$=10$^{-5}$ and $\mu_N$=10$^{-3}$, and population

596 size (a) $N$=25 and (b) $N$=100. Theoretical equilibrium values for $F_{IS}$ at a neutral locus in a

597 self-compatible, random-mating population (dotted black line, from Equation 3, neglecting

598 mutation) and in a self-incompatible population (continuous grey line, from Equation 2) are

599 also represented.

600 **Figure 4** Frequency of deleterious alleles at increasing rates of asexual reproduction under

601 two selection regimes. Box-plots represent median (black line), first and third quartiles (white

602 and grey boxes, for self-compatible and self-incompatible populations respectively) and 5%

603 and 95% percentiles (whiskers) of 1000 observations from the simulations performed at seven

604 rates of asexuality ($c$=0, 0.5, 0.8, 0.9, 0.99, 0.999 and 1), mutation rates $\mu_S$=10$^{-5}$ and $\mu_N$=10$^{-3}$.

605 Population size and viability locus coefficients: (a) $N$=25, $s$=0.1, $h$=0.2, (b) $N$=25, $s$=1,

606 $h$=0.02, (c) $N$=100, $s$=0.1, $h$=0.2, and (d) $N$=100, $s$=1, $h$=0.02.

607 **Figure 5** Mutation load at increasing rates of asexual reproduction under two different

608 selection regimes according to reproductive system. Box-plots represent median (black line),

609 first and third quartiles (box) and 5% and 95% percentiles (whiskers) of 1 000 observations

610 from simulations performed at seven rates of asexual reproduction ($c$=0, 0.5, 0.8, 0.9, 0.99,

611 0.999 and 1) and four population sizes ($N$=25, 50, 100 and 1000). (a) Self-incompatible

612 population with a viability locus with mildly deleterious allele ($s$=0.1, $h$=0.2), (b) self-

613 compatible population with a viability locus with mildly deleterious allele ($s$=0.1, $h$=0.2), (c)

614 self-incompatible population with a viability locus with highly recessive lethal allele ($s$=1,

615 $h$=0.02), and (d) self-compatible population with a viability locus with highly recessive lethal

616 allele ($s$=1, $h$=0.02).

617 **Figure 6** Inbreeding depression at increasing rates of asexual reproduction and under two

618 different selection regimes according to reproductive system. Box-plots represent median

619 (black line), first and third quartiles (box) and 5% and 95% percentiles (whiskers) of 1000

620 observations from simulations performed at seven rates of asexual reproduction ($c$=0, 0.5, 0.8,

621 0.9, 0.99, 0.999 and 1) and four population sizes ($N$=25, 50, 100 and 1000). (a) Self-

622 incompatible population with a viability locus with mildly deleterious allele ($s$=0.1, $h$=0.2),

623 (b) self-compatible population with a viability locus with mildly deleterious allele ($s$=0.1,

624 $h$=0.2), (c) self-incompatible population with a viability locus with highly recessive lethal

625 allele ($s$=1, $h$=0.02), and (d) self-compatible population with a viability locus with highly

626 recessive lethal allele ($s$=1, $h$=0.02).

627    **Figure 7** Fitness ratio between sexually and asexually produced individuals at increasing rates

628    of asexual reproduction under two different selection regimes according to reproductive

629    system. Box-plots represent median (black line), first and third quartiles (box) and 5% and

630    95% percentiles (whiskers) of 1 000 observations from simulations performed at seven rates

631    of asexual reproduction ($c$=0, 0.5, 0.8, 0.9, 0.99, 0.999 and 1) and four population sizes

632    ($N$=25, 50, 100 and 1000). (a) Self-incompatible population with a viability locus with mildly

633    deleterious allele ($s$=0.1, $h$=0.2), (b) self-compatible population with a viability locus with

634    mildly deleterious allele ($s$=0.1, $h$=0.2), (c) self-incompatible population with a viability locus

635    with highly recessive lethal allele ($s$=1, $h$=0.02), and (d) self-compatible population with a

636    viability locus with highly recessive lethal allele ($s$=1, $h$=0.02).

637

638 **Table 1** Genotype fixation probability at the *S*-locus under two different selection regimes at

639 the viability locus

| Population size | Selection regime | Rate of asexual reproduction | | | | | | |
|---|---|---|---|---|---|---|---|---|
| | | $c=0$ | $c=0.5$ | $c=0.8$ | $c=0.9$ | $c=0.99$ | $c=0.999$ | $c=1$ |
| $N=25$ | Mildly deleterious[1] | 0.000 | 0.000 | 0.036 | 0.325 | 0.784 | 0.807 | 0.997 |
| $N=25$ | Lethal recessive[2] | 0.000 | 0.000 | 0.018 | 0.315 | 0.790 | 0.805 | 0.994 |
| $N=50$ | Mildly deleterious | 0.000 | 0.000 | 0.000 | 0.002 | 0.072 | 0.117 | 0.990 |
| $N=50$ | Lethal recessive | 0.000 | 0.000 | 0.000 | 0.001 | 0.075 | 0.090 | 0.989 |
| $N=100$ | Mildly deleterious | 0.000 | 0.000 | 0.000 | 0.000 | 0.001 | 0.001 | 0.983 |

640 The probability of genotype fixation at the *S*-locus (only two alleles in the population) was

641 estimated as the proportion of observations in which this event was recorded. Results in this

642 table are for populations with a mutation rate of $\mu_S=10^{-5}$ at the *S*-locus.

643 [1] Viability locus selection and dominance coefficients: *s*=0.1 and *h*=0.2

644 [2] Viability locus selection and dominance coefficients: *s*=1 and *h*=0.02

645

**Table 2** Fixation probabilities at the viability locus in a self-incompatible population under two different selection regimes.

| Population size | Selection regime | Rate of asexual reproduction | | | | | | |
|---|---|---|---|---|---|---|---|---|
| | | $c$=0 | $c$=0.5 | $c$=0.8 | $c$=0.9 | $c$=0.99 | $c$=0.999 | $c$=1 |
| **(a) Fixation of deleterious allele *a*** | | | | | | | | |
| *N*=25 | Mildly deleterious[1] | 0.066 | 0.049 | 0.056 | 0.050 | 0.085 | 0.081 | 0.067 |
| *N*=50 | Mildly deleterious | 0.000 | 0.000 | 0.001 | 0.000 | 0.001 | 0.001 | 0.001 |
| *N*=100 | Mildly deleterious | 0.000 | 0.000 | 0.000 | 0.000 | 0.000 | 0.000 | 0.000 |
| **(b) Fixation of allele *A*** | | | | | | | | |
| *N*=25 | Mildly deleterious | 0.670 | 0.714 | 0.663 | 0.519 | 0.264 | 0.279 | 0.082 |
| *N*=25 | Lethal recessive[2] | 0.831 | 0.783 | 0.767 | 0.581 | 0.293 | 0.317 | 0.085 |
| *N*=50 | Mildly deleterious | 0.583 | 0.541 | 0.574 | 0.507 | 0.474 | 0.466 | 0.139 |
| *N*=50 | Lethal recessive | 0.661 | 0.635 | 0.596 | 0.564 | 0.518 | 0.493 | 0.150 |
| *N*=100 | Mildly deleterious | 0.314 | 0.306 | 0.320 | 0.294 | 0.274 | 0.290 | 0.187 |
| *N*=100 | Lethal recessive | 0.437 | 0.361 | 0.318 | 0.313 | 0.269 | 0.296 | 0.189 |
| **(c) Fixation of genotype *Aa*** | | | | | | | | |
| *N*=25 | Mildly deleterious | 0.000 | 0.000 | 0.023 | 0.184 | 0.472 | 0.467 | 0.723 |
| *N*=25 | Lethal recessive | 0.000 | 0.000 | 0.013 | 0.212 | 0.551 | 0.569 | 0.879 |
| *N*=50 | Mildly deleterious | 0.000 | 0.000 | 0.000 | 0.000 | 0.026 | 0.050 | 0.585 |
| *N*=50 | Lethal recessive | 0.000 | 0.000 | 0.000 | 0.000 | 0.032 | 0.041 | 0.721 |
| *N*=100 | Mildly deleterious | 0.000 | 0.000 | 0.000 | 0.000 | 0.000 | 0.000 | 0.230 |
| *N*=100 | Lethal recessive | 0.000 | 0.000 | 0.000 | 0.000 | 0.000 | 0.000 | 0.317 |

The probability of fixation at the viability locus (fixation of genotypes, *AA*, *aa* or *Aa*) was estimated as the proportion of observations in which fixation was recorded.

[1] Viability locus selection and dominance coefficients: *s*=0.1 and *h*=0.2

[2] Viability locus selection and dominance coefficients: *s*=1 and *h*=0.02

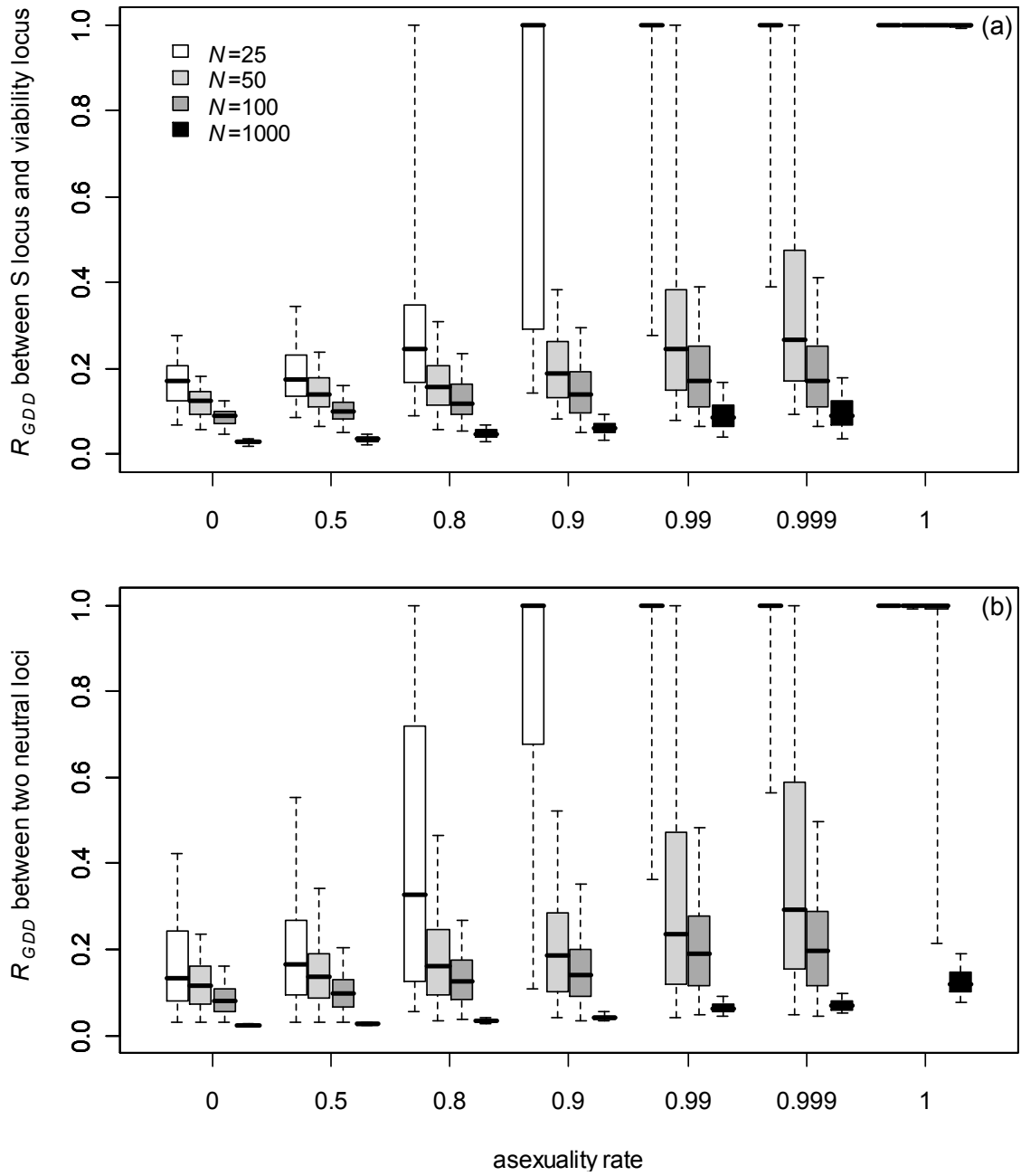

 **Figure 2**

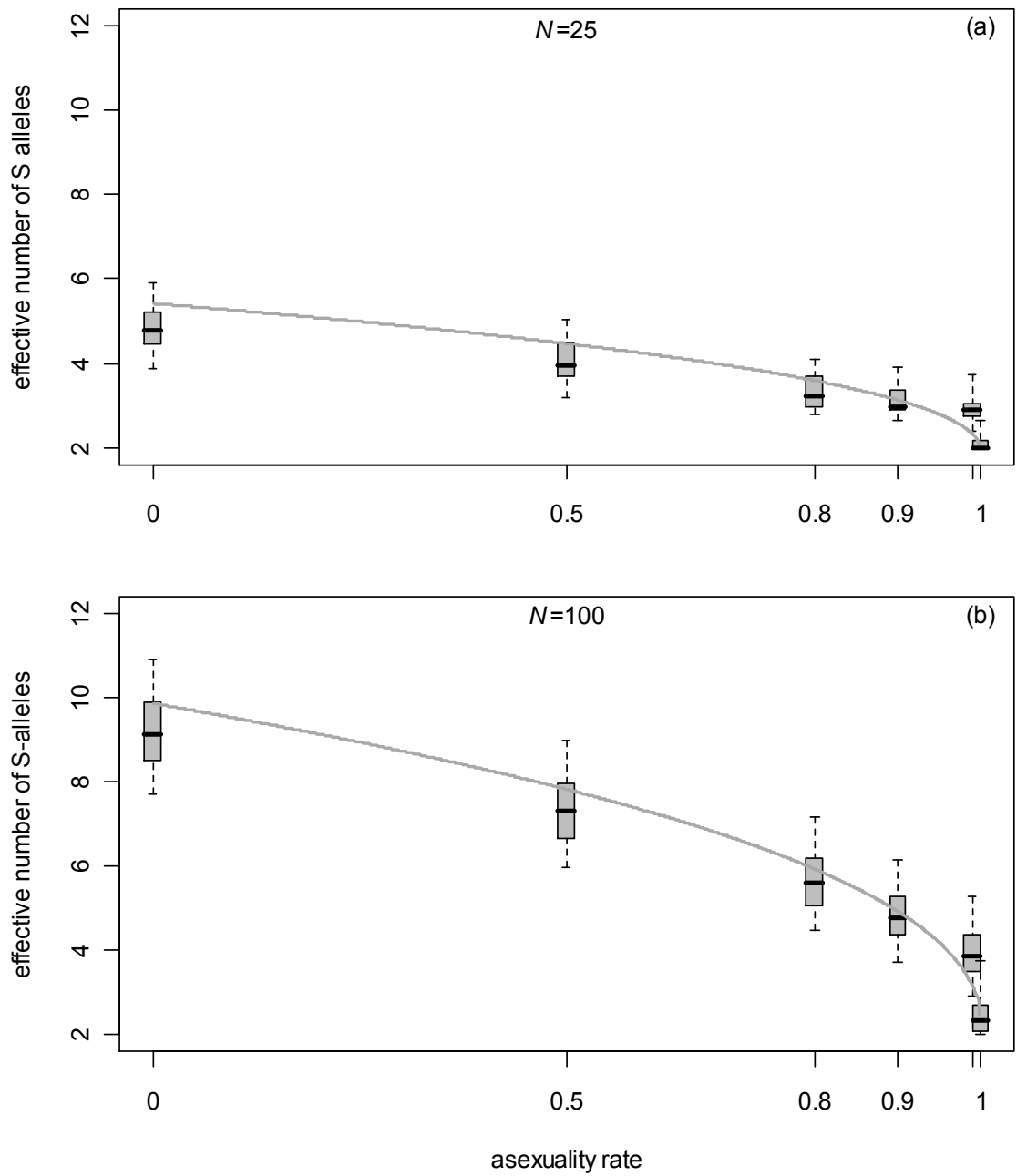

659 **Figure 3**

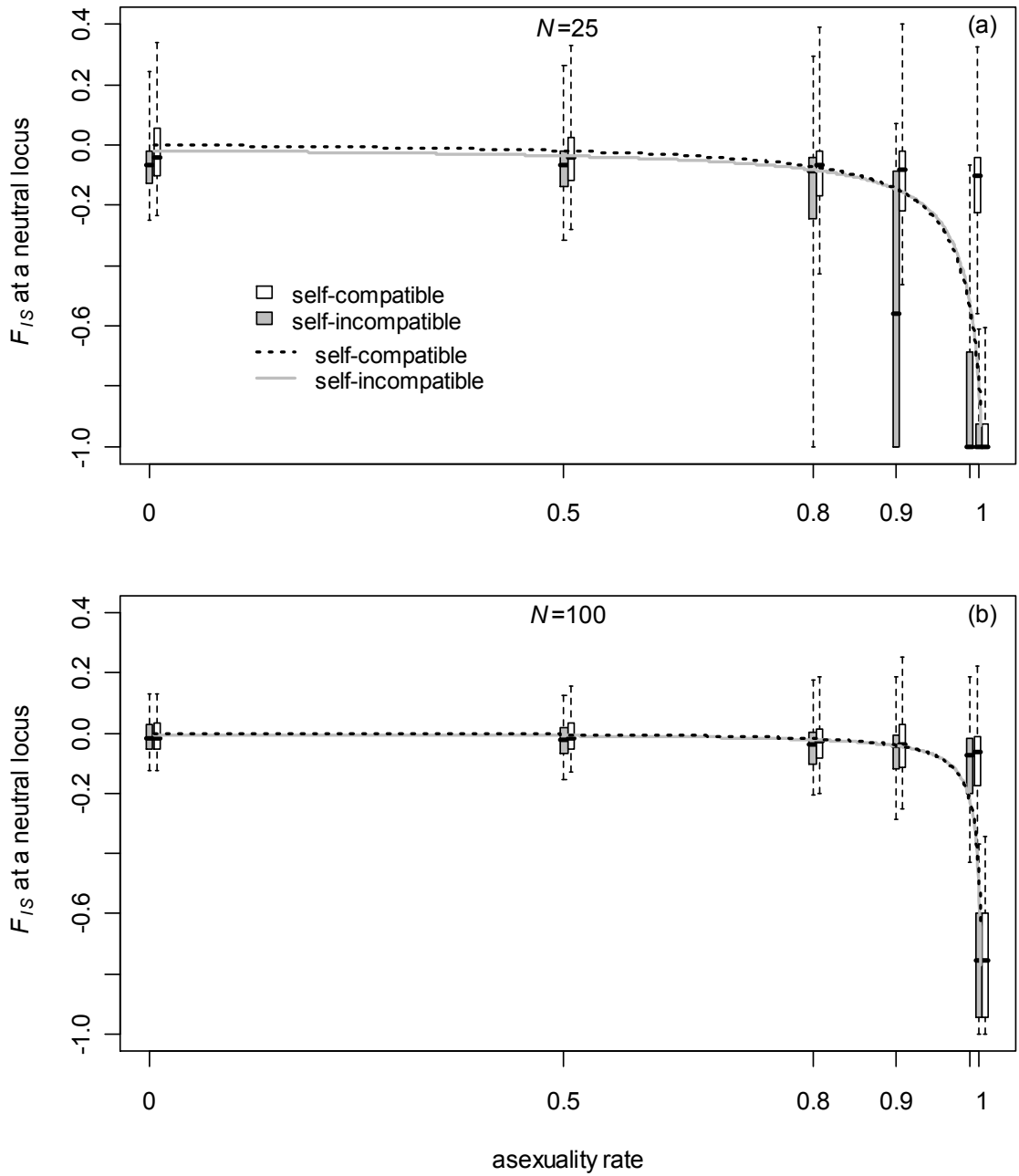

660

661

**Figure 4**

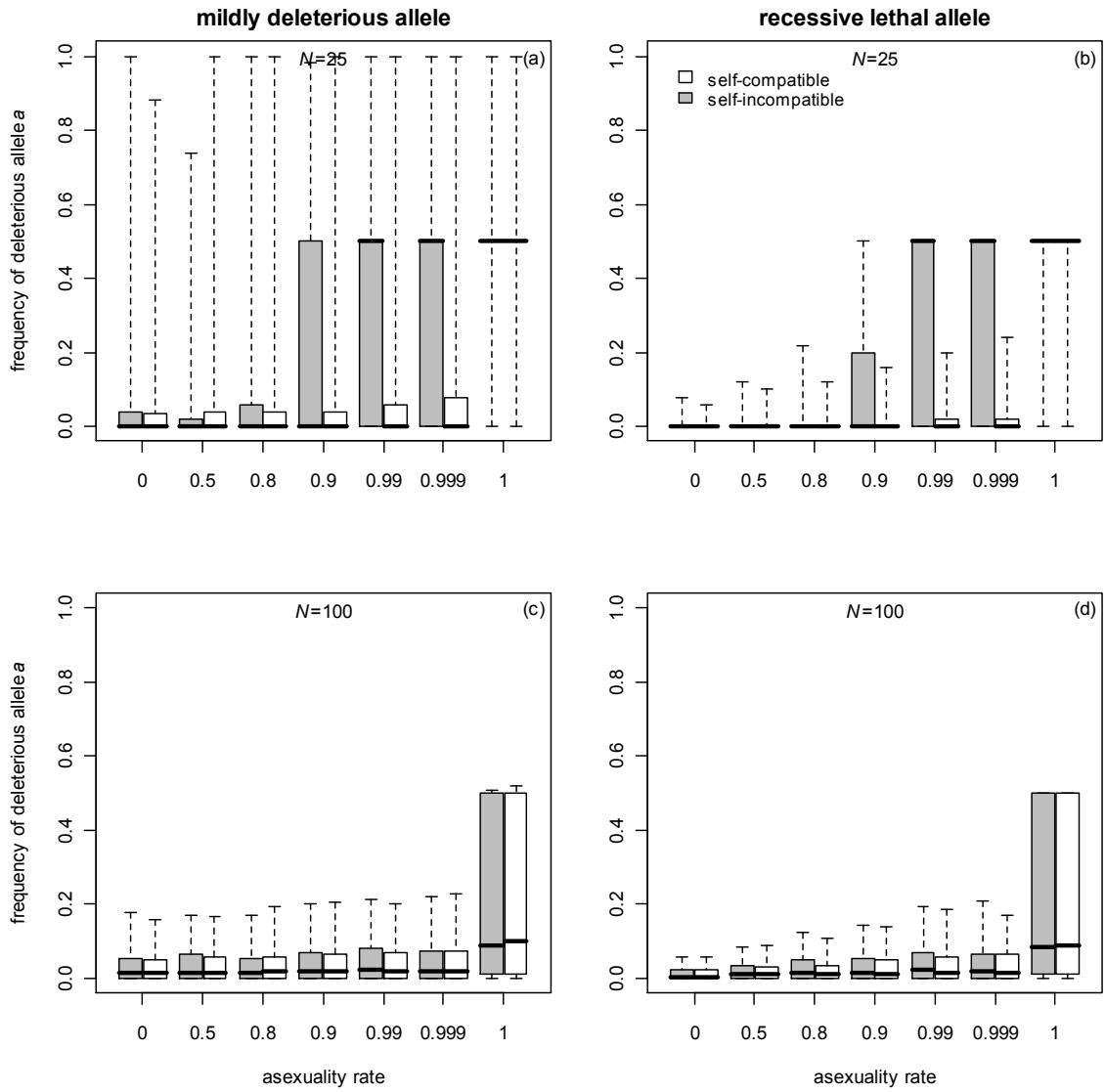

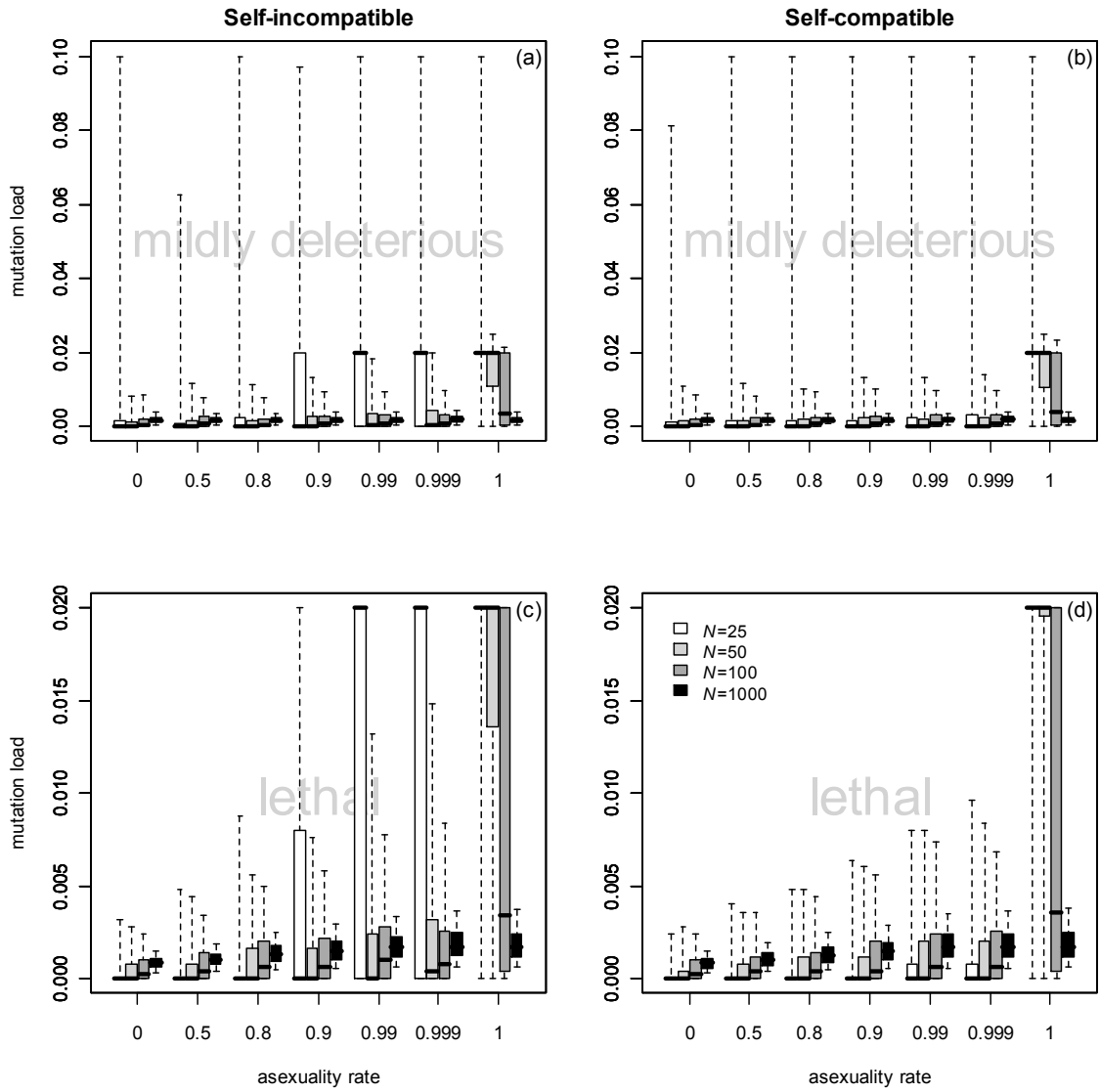

**Figure 6**

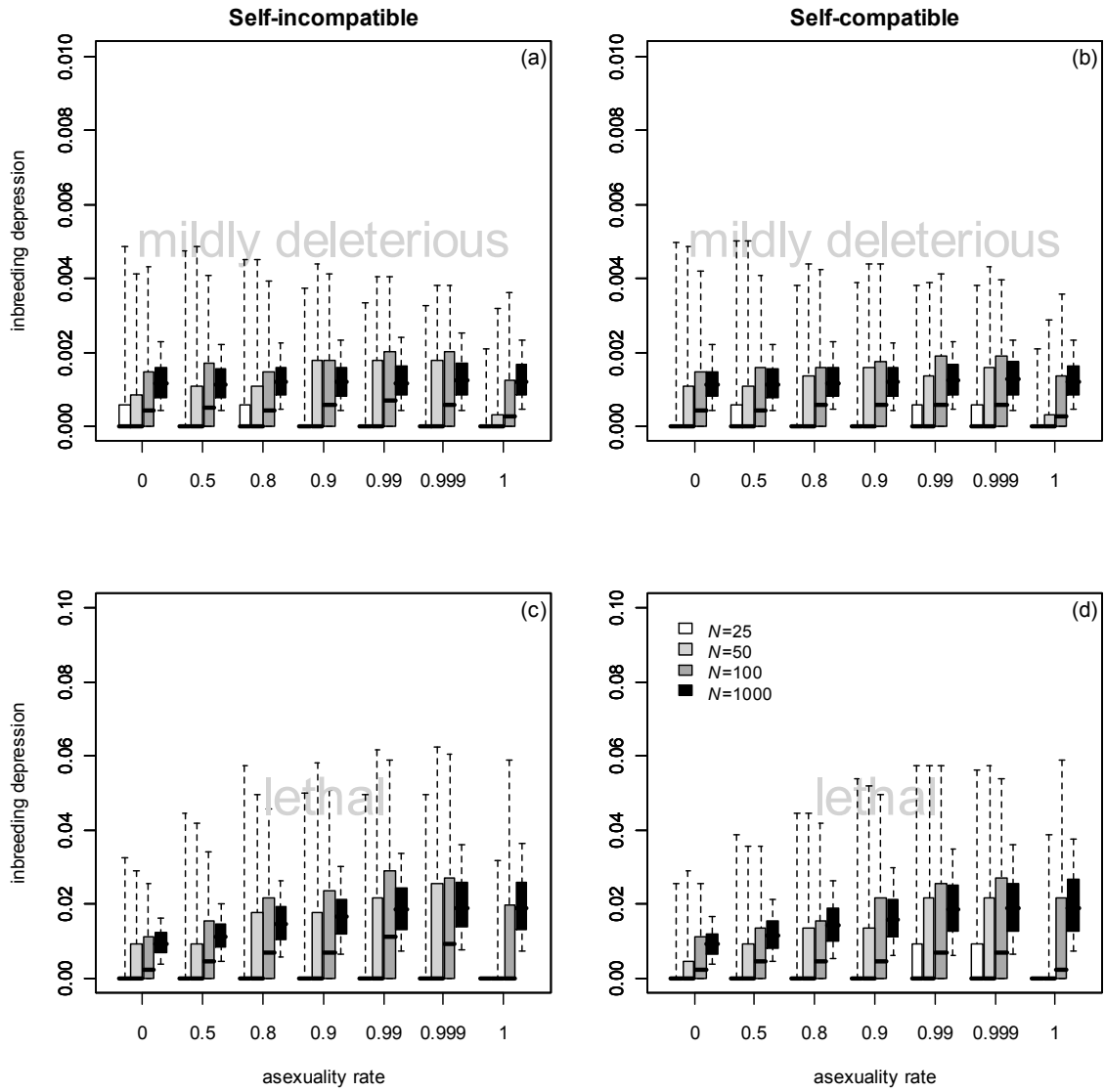

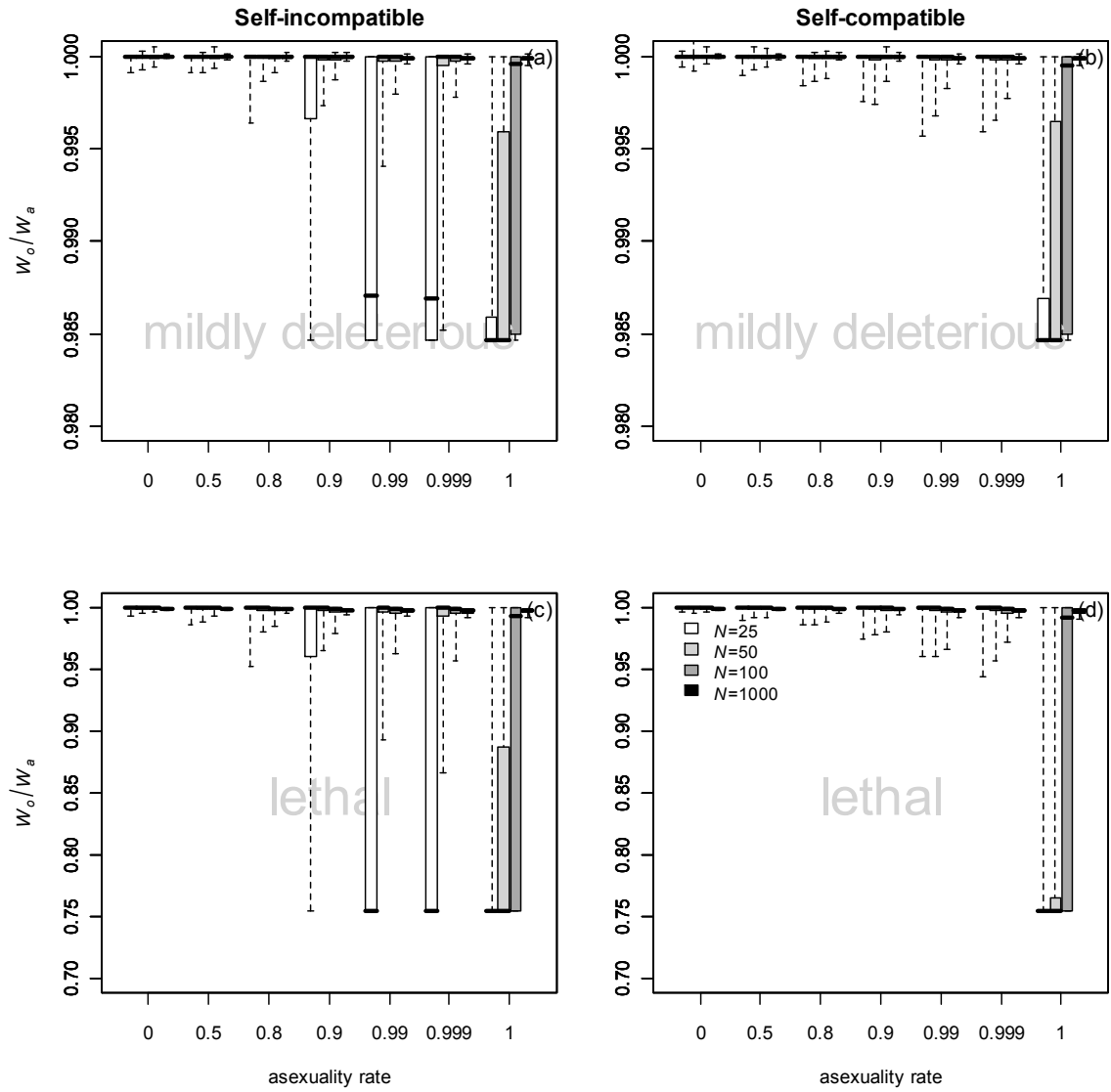